\documentclass[12pt]{article}%
\usepackage{amsmath,latexsym}
\usepackage{graphicx}
\usepackage{amsmath}
\usepackage{amsfonts}
\usepackage{amssymb}%
\begin{document}

\title{{Accounting for some aspects of dark matter and
   dark energy via noncommutative geometry}}
   \author{
Peter K. F. Kuhfittig*\\
\footnote{E-mail: kuhfitti@msoe.edu}
 \small Department of Mathematics, Milwaukee School of
Engineering,\\
\small Milwaukee, Wisconsin 53202-3109, USA}

\date{}
 \maketitle

\begin{abstract}\noindent
The purpose of this paper is to seek a
connection between noncommutative geometry,
an offshoot of string theory, and certain
aspects of dark matter and dark energy.  The
former case is based on a simple mathematical
argument showing that the main manifestation
of dark matter in connection with flat
rotation curves in galaxies and clusters of
galaxies is also a consequence of
noncommutative geometry.  The latter case
requires an examination of the local effect
of noncommutative geometry and the subsequent
extension to the global phenomenon of an
accelerating Universe.\\

\noindent
Keywords\\
Noncommutative Geometry, Dark Matter, Dark Energy\\

\end{abstract}

\section{Introduction}

While it is generally assumed that dark matter
is needed to account for galactic rotation
curves in the outer region of galaxies, it
has already been observed that a
noncommutative-geometry background can
accomplish this goal equally well \cite
{fR12, pK12}.  In the first part of this
paper we seek a mathematical explanation
for this outcome: both dark matter and
noncommutative geometry predict that the
mass inside a sphere of radius $r$ increases
linearly with $r$ in the outward radial
direction.  It follows that dark matter is
not needed to account for flat galactic
rotation curves.  The same is true for
rotation curves in clusters of galaxies.

The second part of this paper makes an
analogous connection between noncommutative
geometry and dark energy.  An analysis of
the local effect of noncommutative geometry
suggests an extension thereof to the global
phenomenon of an accelerating Universe.

The main conclusion is that string theory
in the form of a noncommutative-geometry
background can account for certain aspects
of both dark matter and dark energy.


\section{Noncommutative geometry}

Suppose we start with the general metric of
a static spherically symmetric line element,
using units in which $c=G=1$:
\begin{equation}\label{E:line1}
ds^{2}= -e^{2\Phi(r)}dt^{2}+
\frac{dr^2}{1-\frac{2m(r)}{r}}
+r^{2}(d\theta^{2}+\text{sin}^{2}
\theta\,d\phi^{2});
\end{equation}
here $m(r)$ is the effective mass inside a
sphere of radius $r$ with $m(0)=0$.  We
also require that $\text{lim}_{r\rightarrow
\infty} m(r)/r=0$.

Because of the spherical symmetry, the only
nonzero components of the stress-energy
tensor are $T^t_{\phantom{tt}t}=-\rho(r)$,
the energy density, $T^r_{\phantom{rr}r}=
p_r(r)$, the radial pressure, and
$T^\theta_{\phantom{\theta\theta}\theta}=
T^\phi_{\phantom{\phi\phi}\phi}=p_t(r)$,
the lateral pressure.  The Einstein field
equations can be written in the following
form:
\begin{equation}\label{E:Einstein1}
   \rho(r)=\frac{2m'}{8\pi r^2},
\end{equation}
\begin{equation}\label{E:Einstein2}
   p_r(r)=\frac{1}{8\pi}\left[-\frac{2m}{r^3}
   +\frac{2\Phi'}{r}\left(1-\frac{2m}{r}
   \right)\right],
\end{equation}
and
\begin{equation}\label{E:Einstein3}
   p_t(r)=\frac{1}{8\pi}\left(1-\frac{2m}{r}
   \right)\left[\Phi''-\frac{2m'r-2m}{
   2r(r-2m)}\Phi'+(\Phi')^2+\frac{\Phi'}{r}
   -\frac{2m'r-2m}{2r^2(r-2m)}\right].
\end{equation}
The conservation law $T^{\alpha}
_{\phantom{\beta r}\beta;\,\alpha}=0$ implies
that
\begin{equation}
   p'_r+\Phi'\rho+\Phi'p_r+\frac{2p_r}{r}-
   \frac{2p_t}{r}=0.
\end{equation}
As a result, only Eqs. (\ref{E:Einstein1})
and (\ref{E:Einstein2}) are actually needed,
an observation that will be taken advantage
of later.

Next, we take a brief look at noncommutative
geometry, an area that is based on the
following outcome of string theory:
coordinates may become noncommuting
operators on a $D$-brane \cite {eW96, SW99}.
Here the commutator is
$[\textbf{x}^{\mu},\textbf{x}^{\nu}]=
i\theta^{\mu\nu}$, where $\theta^{\mu\nu}$
is an antisymmetric matrix.  The main idea,
discussed in Refs. \cite{SS1, SS2}, is that
noncommutativity replaces point-like
structures by smeared objects.  (The aim
is to eliminate the divergences that
normally occur in general relativity.)
A natural way to accomplish the smearing
effect is to use a Gaussian distribution
of minimal length $\sqrt{\beta}$ rather
than the Dirac delta function
\cite{NSS06, pK13}.  An equivalent, but
simpler, way is to assume that the energy
density of the static and spherically
symmetric and particle-like gravitational
source has the form \cite {NM08, pKnew}
\begin{equation}\label{E:rho}
  \rho(r)=\frac{M\sqrt{\beta}}
     {\pi^2(r^2+\beta)^2}.
\end{equation}
The point is that the mass $M$ of the particle
is diffused throughout the region of linear
dimension $\sqrt{\beta}$ due to the uncertainty.

To make use of Eq. (\ref{E:rho}), one can keep
the standard form of the Einstein field equations
in the sense that the Einstein tensor retains its
original form but the stress-energy tensor is
modified \cite{NSS06}.  It follows that the
length scale need not be restricted to the
Planck scale.  It is further noted in Ref.
\cite{NSS06} that noncommutative geometry is
an intrinsic property of spacetime and does
not depend on any particular feature such as
curvature.

The gravitational source in Eq. (\ref{E:rho})
results in a smeared mass.  As in Refs.
\cite{SS1, SS2}, the Schwarzschild solution of
the Einstein field equations associated with the
smeared source leads to the line element
\begin{equation}\label{E:line2}
ds^{2}=\\
-\left(1-\frac{2M_{\beta}(r)}{r}\right)dt^{2}
  +\left(1-\frac{2M_{\beta}(r)}{r}\right)^{-1}dr^{2}\\
+r^{2}(d\theta^{2}+\text{sin}^{2}\theta\,d\phi^{2}).
\end{equation}
Here the smeared mass is found to be
\begin{equation}\label{E:M}
   M_{\beta}(r)=\int^r_04\pi (r')^2\rho(r')\,dr'
  \\ = \frac{2M}{\pi}\left(\text{tan}^{-1}
  \frac{r}{\sqrt{\beta}}
  -\frac{r\sqrt{\beta}}{r^2+\beta}\right),
\end{equation}
where $M$ is now the total mass of the source.
Since $\text{lim}_{r\rightarrow 0}M_{\beta}
(r)/r=0$, there is no singularity at $r=0$.

Due to the smearing, the mass of the particle
depends on $\beta$, as well as on $r$.  As in
the case of the Gaussian model, the mass of
the particle is zero at the center and rapidly
increases to $M$.  As a result, from a distance
the smearing is no longer observed and we get
an ordinary particle:
\[
   \text{lim}_{\beta\rightarrow 0}M_{\beta}
   (r)=M.
\]
So the modified Schwarzschild solution becomes
an ordinary Schwarzschild solution in the limit.

\section{The dark-matter hypothesis}

The existence of dark matter was already
hypothesized in the 1930's by Zwicky and
others.  The implications thereof were not
recognized until the 1970's when it was
observed that galaxies exhibit flat
rotation curves (constant velocities)
sufficiently far from the galactic center
\cite{RTF80}.  This observation indicates
that the matter in the galaxy increases
linearly in the outward radial direction.

To recall the reason for this, suppose
$m_1$ is the mass of a star, $v$ its
constant velocity, and $m_2$ the mass of
everything else.  Now multiplying $m_1$ by
the centripetal acceleration yields
\begin{equation}
   m_1\frac{v^2}{r}=m_1m_2\frac{G}{r^2},
\end{equation}
where $G$ is Newton's gravitational
constant.  Using geometrized units
($G=c=1$), we obtain the linear form
\begin{equation}\label{E:linear1}
   m_2=rv^2,
\end{equation}
as asserted.  Eq. (\ref{E:linear1})
essentially characterizes the dark-matter
hypothesis.

Consider next a thin spherical shell of radius
$r=r_0$.  So instead of a smeared object located
at the origin, we now have a smeared spherical
surface.  We consider the smearing in the
outward radial direction only, since that is
the analogue of the smeared particle at the
origin.  The energy density in Eq. (\ref{E:rho})
must therefore be replaced by
\begin{equation}
    \rho(r)=\frac{M_{r_0}\sqrt{\beta}}
    {\pi^2[(r-r_0)^2+\beta]^2},
\end{equation}
which is simply a translation in the
$r$-direction. Then the smeared mass of the
shell becomes
\begin{equation}\label{E:m(r)}
   m_{\beta}(r-r_0)=\\ \frac{2M_{r_0}}{\pi}
   \left[\text{tan}^{-1}\frac{r-r_0}
   {\sqrt{\beta}}-\frac{(r-r_0)\sqrt{\beta}}
   {(r-r_0)^2+\beta}\right].
\end{equation}
Observe that, analogously,
\[
   \text{lim}_{\beta\rightarrow 0}\,m_{\beta}
   (r-r_0)=M_{r_0}.
\]
So the mass of the shell is zero at $r=r_0$
and rapidly rises to $M_{r_0}$.

Since we are moving in the outward radial
direction, we can replace ``mass of the shell"
by ``change in mass per unit length in the
$r$-direction," still to be denoted by
$m_{\beta}(r-r_0)$ in Eq. (\ref{E:m(r)}).
Alternatively, if $M_T(r-r_0)$ is the total mass,
then the mass of the shell of thickness $dr$
(i.e., the change in $M_T(r-r_0)$) becomes the
differential
\[
   dM_T(r-r_0)=\frac{dM_T(r-r_0)}{dr}dr=
      m_{\beta}(r-r_0)dr.
\]
Either way, $m_{\beta}(r-r_0)$ and $M_{r_0}$ in
Eq. (\ref{E:m(r)})
are dimensionless in our geometrized units.  The
total smeared mass is therefore given by
\begin{multline}\label{E:total}
  M_T(r-r_0)=\int^{r-r_0}_0 m_{\beta}(r')dr'=
  \frac{2M_{r_0}}{\pi}\left[(r-r_0)\text{tan}^{-1}
  \frac{r-r_0}{\sqrt{\beta}}-\sqrt{\beta}
  \,\,\text{ln}\,[(r-r_0)^2+\beta]\right]\\
  =\frac{2M_{r_0}}{\pi}(r-r_0)\left[\text{tan}^{-1}
  \frac{r-r_0}{\sqrt{\beta}}-\sqrt{\beta}
  \,\frac{\text{ln}\,[(r-r_0)^2+\beta]}
  {r-r_0}\right].
\end{multline}
(So $M_T(r-r_0)$ has units of length.)  For the
expression inside the brackets, we have
\[
  \text{lim}_{r\rightarrow \infty}
  \left[\text{tan}^{-1}
  \frac{r-r_0}{\sqrt{\beta}}-\sqrt{\beta}
  \,\frac{\text{ln}\,[(r-r_0)^2+\beta]}
  {r-r_0}\right]=\frac{\pi}{2}-0.
\]
It follows that $M_T(r-r_0)$ has the linear
form
\begin{equation}\label{E:linear2}
   M_T(r-r_0)=M_{r_0}(r-r_0).
\end{equation}
(We could simply say that $M_T(r)=M_{r_0}r$,
in agreement with Eq. (\ref{E:linear1}).)  Not
only does this provide an alternative to the
dark-matter hypothesis, the geometric
interpretation of the gravitational pull due
to dark matter is very much in the spirit of
Einstein's theory, which replaces the concept
of \emph{gravitational force} by the
geometric concept of \emph{curvature}.

To connect Eq. (\ref{E:linear2}) to the
tangential velocity, we return to Eq.
(\ref{E:total}) and observe that
\begin{equation*}
\frac{2M_{r_0}}{\pi}\left[(r-r_0)\text{tan}^{-1}
  \frac{r-r_0}{\sqrt{\beta}}-\sqrt{\beta}
  \,\,\text{ln}\,[(r-r_0)^2+\beta]\right]
  \rightarrow M_{r_0}(r-r_0)
\end{equation*}
as $r\rightarrow\infty$ or $\beta
\rightarrow 0$, or both.  Here $\beta$ is
necessarily small to start with; so if $r$
is reasonably large, we have
\[
  M_{T}(r)\approx M_{r_0}(r-r_0).
\]
So from Eq. (\ref{E:linear1}), $v^2r\approx
M_{r_0}(r-r_0)$ and
\[
   v^2\approx M_{r_0}\left(1-\frac{r_0}{r}
   \right).
\]
We conclude that $v^2$ is approximately
equal to the change in the smeared mass
per unit length.

A similar conclusion can be reached from
galaxy cluster observations.  According to
the modified gravity model, the total mass
of the cluster is the sum of the baryonic
mass, consisting mainly of intracluster
gas and the part that is attributable to
the modified gravity or to dark matter.
The baryonic mass has a density given by
\cite{CF76, RB02}
\begin{equation}
   \rho(r)=\rho_0\left(1+\frac{r^2}{r_c^2}
   \right)^{-3\beta/2},
\end{equation}
where $r_c$ is the core radius and $\beta$
and $\rho_0$ are constants.  According to
Lobo \cite{fL09}, the total mass is
\begin{equation}
   M(r)=\frac{3k_B\beta T}{\mu m_pG}
   \frac{r^3}{r_c^2+r^2},
\end{equation}
where $k_B$ is Boltzmann's constant, $T$
is the gas temperature, $\mu\approx 0.61$
is the mean atomic weight of the gas
particles, and $m_p$ is the proton mass.
Observe that for large $r$, $M(r)$ has
the same linear form obtained above.

\emph{Remark:} It is important to note that
we are examining only one aspect of dark
matter, accounting for the flat rotation
curves.  Evidence for dark matter also
comes from other sources, such as the need
to explain the CMB temperature anisotropy,
which is beyond the scope of this study.
On the other hand, it is shown in Ref.
\cite{CCM15} that dark matter also emerges
from noncommutative geometry in a more
general cosmological setting, as our
extension to galaxy clusters has
confirmed.

\section{Dark energy}\label{S:dark}

A major discovery in the late 1990's was that
our Universe is undergoing an accelerated
expansion \cite{aR98, sP99}, i.e.,
$\overset{..}{a}>0$ in the Friedmann
equation
\begin{equation}\label{E:Friedmann}
   \frac{\overset{..}{a}(t)}{a(t)}
   =-\frac{4\pi}{3}(\rho+3p).
\end{equation}
Here $p=p_r=p_t$ since in a cosmological
setting we are dealing with a homogeneous
distribution of matter.  The acceleration
is caused by a negative pressure \emph{dark
energy}.  In particular, if the equation of
state is $p=\omega\rho$, then a value of
$\omega <-1/3$ is required for an
accelerated expansion.  (Current data favor
$\omega =-1$, which is equivalent to assuming
Einstein's cosmological constant
\cite{rB12}.)

To make use of
\begin{equation*}
  \rho(r)=\frac{M\sqrt{\beta}}
     {\pi^2(r^2+\beta)^2}
\end{equation*}
in the cosmological model (\ref{E:Friedmann}),
we need to recall that our Universe is a
3-sphere, having neither a center nor an edge.
So any point can be chosen for the origin of
the above $\rho(r)$.  Moreover, the scale
factor $a(t)$ in the FLRW model
\begin{equation}\label{E:line3}
  ds^{2}= -dt^{2}+a(t)\left[\frac{dr^2}{1-Kr^2}
  +r^{2}(d\theta^{2}+\text{sin}^{2}
  \theta\,d\phi^{2})\right]
\end{equation}
refers to Eq. (\ref{E:Friedmann}).  Eq.
(\ref{E:line3}) now suggests that
$\Phi(r)\equiv 0$ for the function $\Phi(r)$
in Eq. (\ref{E:line1}).  So for an arbitrarily
chosen particle, the line element becomes
\begin{equation}\label{E:line4}
ds^{2}= -dt^{2}+
\frac{dr^2}{1-\frac{2m(r)}{r}}
+r^{2}(d\theta^{2}+\text{sin}^{2}
\theta\,d\phi^{2}),
\end{equation}
where $m(r)=M_{\beta}(r)$ in Eq. (\ref{E:M}).

Recalling that $\Phi \equiv 0$, Eq.
(\ref{E:Einstein2}) now yields
\begin{equation}\label{E:local}
   \rho +3p=\frac{M\sqrt{\beta}}
     {\pi^2(r^2+\beta)^2}+\frac{3}{8\pi}
     \left(-\frac{1}{r^3}\right)
     \frac{4M}{\pi}\left(\text{tan}^{-1}
  \frac{r}{\sqrt{\beta}}
  -\frac{r\sqrt{\beta}}{r^2+\beta}\right),
\end{equation}
but only near the origin.

Before trying to generalize this result to a
cosmological setting, let us examine $\rho +3p$
in the neighborhood of the origin by letting
$r=a\sqrt{\beta}$, $a>0$.  Eq. (\ref{E:local})
now yields
\begin{equation}\label{E:acceleration}
   -\frac{4\pi}{3}(\rho +3p)=-\frac{4\pi}{3}
   \frac{M}{\pi^2}\frac{1}{\beta^{3/2}}
   \left[\frac{1}{(a^2+1)^2}-\frac{3}{2a^3}
   \left(\text{tan}^{-1}a-\frac{a}{a^2+1}
   \right)\right].
\end{equation}
The result can best be seen qualitatively by
plotting $\rho +3p$ against $a$, as shown in
Fig. 1.  So $\rho +3p$
\begin{figure}[tbp]
\begin{center}
\includegraphics[width=0.8\textwidth]{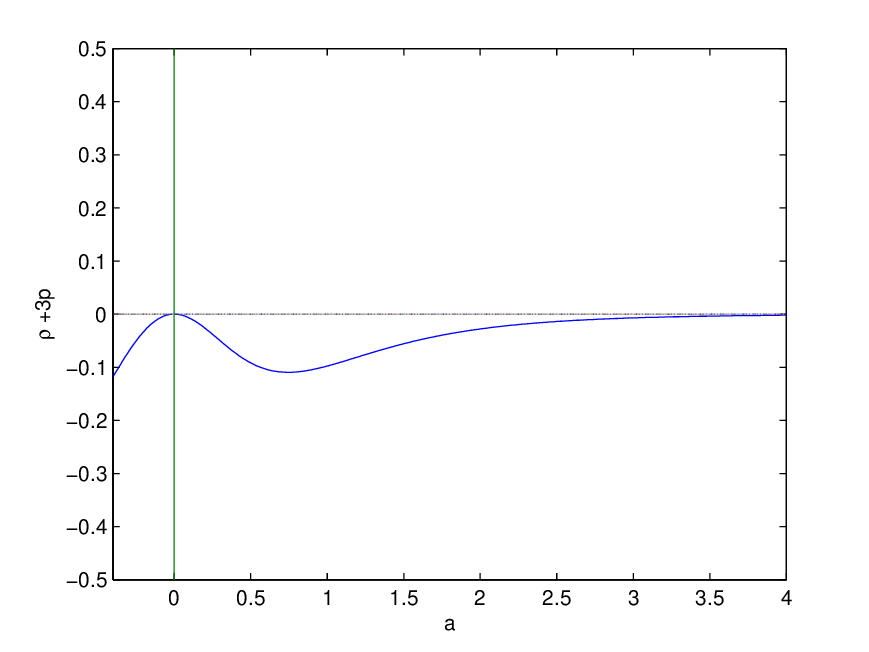}
\end{center}
\caption{$\rho +3p$ is plotted against $a$.}
\end{figure}
is zero at the origin, then becomes negative,
before approaching zero asymptotically.  The
asymptotic behavior is to be expected since
from a distance, the smearing is no longer
apparent.

We can now assert that
\begin{equation}\label{E:dark}
   -\frac{4\pi}{3}(\rho +3p)>0
\end{equation}
in the neighborhood of every particle.
Moreover, we are now in a vacuum, which is
teeming with virtual particles.  These
particles are extremely short lived, but
given that all particles have a finite
lifetime, it is generally assumed that
there is no absolute distinction between
virtual and ordinary particles.  Inequality
(\ref{E:dark}) would therefore hold in
the vicinity of every point.  The
cumulative effect would therefore be an
accelerated expansion on a cosmological
scale.

Inequality (\ref{E:dark}) is consistent
with the discussion in Ref. \cite{NSS06},
which asserts that the collapse of a smeared
particle to a point mass is prevented by a
kind of ``quantum pressure," an outward push
induced by noncommuting coordinate quantum
fluctuations.

\section{Conclusion}
This paper discusses certain manifestations of
string theory in the form of noncommutative
geometry.  The first part of this paper examines
a particular aspect of dark matter, accounting
for flat galactic rotation curves.  It is
shown that a noncommutative-geometry background
agrees with the dark-matter assumption, whose
basic manifestation is the linearly increasing
mass in the outward radial direction.
Connections to other aspects of dark matter,
such as structure formation, are thereby left
open.  However, according to Ref. \cite{CCM15},
dark matter does emerge from noncommutative
geometry in a cosmological setting, as
exemplified by our extension to clusters of
galaxies.

The second part of this paper discusses a
particular aspect of dark energy by examining
the local effect of the smearing that
characterizes noncommutative geometry.  Since
the Universe is a 3-sphere, any point can serve
as the origin for $\rho(r)$.  As discussed at
the end of Sec. \ref{S:dark}, in a vacuum, the
existence of virtual particles everywhere then
implies that $(-4\pi/3)(\rho +3p)>0$ in the
neighborhood of the origin and hence of every
point.  The cumulative effect is an acceleration
on a cosmological scale.  So if the virtual
particles are indeed the cause of the
acceleration, then it is appropriate to say
that dark energy can be viewed as vacuum
energy.

Concluding comment: accounting for both dark
matter and dark energy may be considered a
promissing step toward obtaining empirical
evidence for string theory, given that
noncommutative geometry is an offshoot
thereof.

\end{document}